\documentclass[aps,pre,twocolumn,superscriptaddress,showkeys,amsmath,amssymb,showpacs,floatfix]{revtex4-1}
\usepackage{graphicx}
\usepackage{dcolumn}
\usepackage{bm}
\usepackage{epsfig}
\usepackage{longtable}
\usepackage{subfigure}
\usepackage{multirow}
\usepackage{enumerate}
\usepackage{url}
\usepackage{afterpage}
\usepackage{xcolor}
\usepackage[normalem]{ulem}  
\usepackage[colorlinks=true,linkcolor=blue,citecolor=blue,urlcolor=blue]{hyperref}  

\allowdisplaybreaks

\definecolor{darkgreen}{rgb}{0,0.5,0}

\begin{document}
\newcommand{\kpe}{\mathbf{k}\!\cdot\!\mathbf{p}\,}
\newcommand{\Kpe}{\mathbf{K}\!\cdot\!\mathbf{p}\,}
\newcommand{\bfr}{ {\bf r}} 
\newcommand{\bfrp}{ {\bf r'}} 
\newcommand{\tp}{ {t^\prime}} 
\newcommand{\bfrpp}{ {\bf r^{\prime\prime}}} 
\newcommand{\bfR}{ {\bf R}} 
\newcommand{\bfRp}{ {\bf R'}} 
\newcommand{\bfG}{ {\bf G}} 
\newcommand{\bfGp}{ {\bf G'}} 
\newcommand{\bfzero}{ {\bf 0}} 
\newcommand{\bfq}{ {\bf q}} 
\newcommand{\bfp}{ {\bf p}} 
\newcommand{\bfk}{ {\bf k}} 
\newcommand{\dv}{ {\Delta \hat{v}}} 
\newcommand{\sigmap}{\sigma^\prime} 
\newcommand{\omegap}{\omega^\prime} 
\newcommand{\omegapp}{\omega^{\prime\prime}} 
\newcommand{\bracketm}[1]{\ensuremath{\langle #1   \rangle}}
\newcommand{\bracketw}[2]{\ensuremath{\langle #1 | #2  \rangle}}
\newcommand{\bracket}[3]{\ensuremath{\langle #1 | #2 | #3 \rangle}}
\newcommand{\ket}[1]{\ensuremath{| #1 \rangle}}
\newcommand{\GnWn}{\ensuremath{G_\text{0}W_\text{0}}\,}
\newcommand{\tn}[1]{\textnormal{#1}}
\newcommand{\f}[1]{\footnotemark[#1]}
\newcommand{\mc}[2]{\multicolumn{1}{#1}{#2}}
\newcommand{\mcs}[3]{\multicolumn{#1}{#2}{#3}}
\newcommand{\mcc}[1]{\multicolumn{1}{c}{#1}}
\newcommand{\refeq}[1]{(\ref{#1})} 
\newcommand{\refcite}[1]{Ref.~\cite{#1}} 
\newcommand{\refsec}[1]{Sec.~\ref{#1}} 
\newcommand\opd{d}
\newcommand\im{i}
\def\bra#1{\mathinner{\langle{#1}|}}
\def\ket#1{\mathinner{|{#1}\rangle}}
\newcommand{\braket}[2]{\langle #1|#2\rangle}
\def\Bra#1{\left<#1\right|}
\def\Ket#1{\left|#1\right>}
\def\onlinecite{\cite[left=,right=]}

\newcommand{\fatr}{\mathbf{r}}

\title[]{All-electron Quasiparticle Self-consistent \texorpdfstring{$GW$}{GW} for Molecules and Periodic Systems within the Numerical Atomic Orbital Framework}
\date{\today}

\author{Bohan Jia}
\affiliation{Institute of Physics, Chinese Academy of Sciences, Beijing 100190, China}
\affiliation{School of Physical Sciences, University of Chinese Academy of Sciences, Beijing 100049, China}
\author{Min-Ye Zhang}
\email{minyez@iphy.ac.cn}
\affiliation{Institute of Physics, Chinese Academy of Sciences,  Beijing 100190, China}
\affiliation{The NOMAD Laboratory at the Fritz Haber Institute of the Max Planck Society, Berlin 14195, Germany}
\author{Ziqing Guan}
\affiliation{Institute of Physics, Chinese Academy of Sciences,  Beijing 100190, China}
\affiliation{School of Physical Sciences, University of Chinese Academy of Sciences, Beijing 100049, China}
\author{Huanjing Gong}
\affiliation{Institute of Physics, Chinese Academy of Sciences,  Beijing 100190, China}
\affiliation{School of Physical Sciences, University of Chinese Academy of Sciences, Beijing 100049, China}
\author{Xinguo Ren\thanks{Corresponding author: renxg@iphy.ac.cn}}
\email{renxg@iphy.ac.cn}
\affiliation{Institute of Physics, Chinese Academy of Sciences, Beijing 100190, China}

\begin{abstract}
	We report an all-electron implementation of the quasiparticle self-consistent $GW$ (QS$GW$) method for molecular and periodic systems within the framework of numerical atomic orbitals (NAOs), as implemented in the LibRPA software package. Our implementation is based on the space-time formalism, combined with the localized resolution-of-identity approximation to treat two-electron quantities. We found that analytical continuation of the self-energy matrix, in combination with the ``Mode B" QS$GW$ scheme, can yield stable self-consistent quasiparticle energy spectra. 
	Systematic benchmark calculations on molecules and crystalline solids (including typical semiconductors and wide-gap insulators) demonstrate that our NAO-based QS$GW$ scheme yields molecular ionization potentials and quasiparticle band gaps for periodic solids that are consistent with reference results from established implementations. Our work opens the way for large-scale QS$GW$ calculations, taking advantage of the NAO-based low-scaling algorithm previously developed for the  $G^0W^0$ method. 
\end{abstract}

\maketitle

\raggedbottom

\section{\label{sec:introduction}Introduction}
In computational chemistry and condensed matter physics, accurately solving the quantum many-body problem from first principles remains a fundamental challenge.
While Kohn-Sham (KS) density functional theory (DFT) \cite{Hohenberg/Kohn:1964,Kohn/Sham:1965} is the most widely adopted approach, it suffers from well-known limitations of approximate exchange-correlation (XC) functionals, particularly in predicting excited-state properties and describing strongly correlated systems.
For a large variety of molecules and materials, many-body perturbation theory within the $GW$ approximation \cite{Hedin:1965,Strinati/Mattausch/Hanke:1980,Hybertsen/Louie:1987,Godby/etal:1988,Aryasetiawan/Gunnarsson:1995,Aulbur/Jonsson/Wilkins:2000,Onida/Reining/Rubio:2002,Rinke/etal:2005,Marom/etal:2012,Golze/Dvorak/Rinke:2019,Reining2018,Lange2018} offers a powerful approach for quantitatively describing single-particle excitations. Within $GW$, the many-body self-energy is given by the product of the Green's function $G$ and the screened Coulomb interaction $W$, which captures the dominant screening effect in a many-body system and yields physically meaningful quasiparticle energies.

Over the past four decades, the $GW$ method has been developed into a versatile approach encompassing a variety of practical computational schemes, depending on whether and how self-consistency is incorporated. The simplest variant of $GW$ is the so-called $G^0W^0$ scheme, where the Green's function $G^0$ and screened Coulomb interaction $W^0$ are constructed from the outputs of a preceding mean-field calculation.
$G^0W^0$ based on KS-DFT starting points with local or semilocal exchange-correlation functionals, such as the local density approximation (LDA) or generalized gradient approximations (GGAs) \cite{Perdew/Burke/Ernzerhof:1996,Hybertsen/Louie:1987,Onida/Reining/Rubio:2002,Reining2018}, has proven highly successful in predicting quantitatively accurate electronic band gaps for a large variety of materials.

However, as a one-shot perturbative scheme, $G^0W^0$ results inevitably exhibit a noticeable dependence on the starting point, i.e., the XC functional employed in the preceding mean-field calculation \cite{Bruneval2013,Korzdorfer2012,Chen2014,Gant2022,Zhang2022_Orbitals}. In particular, for materials with complex electronic structures where LDA and GGAs perform poorly, it is often not
\textit{a priori} clear which starting point is appropriate. Furthermore, many applications require not only quasiparticle energies but also updated charge densities and wavefunctions beyond KS-DFT, which $G^0W^0$ cannot provide.

In principle, solving the set of $GW$ equations in a fully self-consistent manner (the so-called sc$GW$ scheme) is feasible \cite{Ku/Eguiluz:2002,Stan/Dahlen/Leeuwen:2006,Rostgaard/Jacobsen/Thygesen:2010,Caruso2013,Caruso/etal:2012,Kutepov:2016,Kutepov:2017,Shishkin2007b,grumet2018beyond,Holm1998,Koval2014,wen2024comparing,Maggio2017,Kutepov2009}.
In the fully self-consistent limit, sc$GW$ is conserving in the sense of Baym and Kadanoff and removes the explicit starting-point dependence \cite{BaymKadanoff1961,Baym1962}.
However, sc$GW$ is computationally demanding and, without vertex corrections, often overestimates band gaps for semiconductors and insulators \cite{Shishkin2007b,shishkin2007accurate}.
Moreover, neither $G^0W^0$ nor sc$GW$ provides an updated single-particle Hamiltonian with explicit quasiparticle wavefunctions, which limits their utility in theoretical analyses that require orbital information, such as the investigation of topological materials.
These limitations motivate the development of an alternative approach that can deliver quasiparticle solutions while maintaining theoretical rigor. The quasiparticle self-consistent $GW$ (QS$GW$) method, developed by
Faleev, van Schilfgaarde, and Kotani \cite{Faleev2004PRL,VanSchilfgaarde2006,Kotani2007a}, fulfills these requirements.

The fundamental principle of QS$GW$ is to approximate the dynamic self-energy $\Sigma(\omega)$ by a static, Hermitian XC potential $V^{\mathrm{xc}}$.
This potential is used to iteratively update a static mean-field Hamiltonian until self-consistency is reached, enabling the simultaneous optimization of quasiparticle energies and wavefunctions.
Different prescriptions exist for constructing the static, Hermitian $V^{\mathrm{xc}}$ from $\Sigma(\omega)$, including the commonly used ``Mode A''/``Mode B'' constructions \cite{VanSchilfgaarde2006} and similarity renormalization group (SRG)-based approaches \cite{Marie2023}.
Moreover, QS$GW$ can be systematically improved by incorporating vertex corrections (e.g., ladder diagrams in $W$) \cite{Cunningham2023,Kutepov2022,Forster2025,sakakibara2020finite,Forster2022_W} and lattice-polarization effects on $W$ \cite{lambrecht2017}.
In this work, we adopt the standard formulation proposed by the original authors \cite{Faleev2004PRL,VanSchilfgaarde2006,Kotani2007a}.
The reliability and broad applicability of this approach have been extensively validated across a diverse range of materials, from conventional semiconductors to complex transition metal oxides \cite{Das2015,salas2022electronic}.

Naturally, the computational efficiency of the QS$GW$ method is closely tied to the underlying numerical implementation.
Over the years, QS$GW$ has been implemented within various numerical frameworks, including linear muffin-tin orbitals (LMTO)~\cite{Kotani2007a}, augmented-plane-wave based all-electron methods~\cite{deguchi2016accurate,salas2022electronic}, and Gaussian-type orbitals (GTOs)~\cite{Lei2022}. More recently, implementations on GPU architectures have also been reported to accelerate the QS$GW$ workflow~\cite{obata2025efficient}.

In recent years, numerical atomic orbitals (NAOs) have emerged as a powerful basis-set framework for electronic structure calculations.
NAOs offer several key advantages:
First, the flexible shape of NAO basis functions naturally facilitates the description of localized states---such as core electrons and strongly correlated $d/f$ electrons---enabling all-electron calculations that are free of pseudopotential approximations.
Second, the compactness of NAOs significantly reduces the dimensionality of the Hilbert space. Unlike plane-wave methods, which require thousands of basis functions and a large number of unoccupied states to converge the screened Coulomb interaction and the self-energy, NAOs typically require only tens to hundreds of functions, thereby greatly alleviating the slow convergence with respect to the number of unoccupied states. Furthermore, when combined with the resolution of identity (RI) technique \cite{feyereisen1993use,vahtras1993integral,weigend1998ri,Forster2020}, two-electron integrals within the NAO framework can be efficiently decomposed, drastically reducing memory requirements and improving computational scaling.
Benchmarks using established NAO codes, such as FHI-aims, demonstrate that the accuracy achieved on the GW100 test set \cite{vanSetten2015GW100} is comparable to that of plane-wave implementations, while the compactness of NAO basis sets and low-scaling formulations can offer favorable efficiency \cite{ren2012resolution,Forster2020,Ren2021PRM,shi2024subquadratic}.

Despite the existence of molecular and periodic $G^0W^0$ implementations based on NAOs \cite{ren2012resolution,Ren2021PRM,Zhang2026arXiv},
an all-electron QS$GW$ implementation within the NAO framework has not yet been reported.
The absence of the QS$GW$ feature has hindered the application of NAO-based methods to complex systems where self-consistency is essential, such as topological materials and strongly correlated oxides.
Developing such an implementation is therefore highly desirable, as it would combine the computational efficiency of NAOs with the rigorous quasiparticle framework of QS$GW$.

In this paper, we present an implementation of the QS$GW$ method based on all-electron NAOs within the LibRPA software package, applicable to both molecules and periodic systems.
We systematically benchmark the performance of our implementation for the vertical ionization potentials (IPs) of small molecules from the GW100 test set, as well as the band gaps of a diverse set of semiconductors and insulators.
Furthermore, we present a detailed comparison with the $G^0W^0$ method and other QS$GW$ implementations to comprehensively evaluate the accuracy and reliability of our approach.

This paper is organized as follows.
In Sec.~\ref{sec:Theoretical_Framework}, we outline the theoretical framework, detailing the QS$GW$ formalism and its implementation using the NAO-based real-space formalism combined with the localized resolution of identity (LRI) technique.
Sec.~\ref{sec:implementation} discusses specific implementation details in LibRPA, including the self-consistency workflow, the convergence behavior, and an analysis of the basis dependence in the analytic continuation process.
Sec.~\ref{sec:results} presents benchmark results for molecular systems (Sec.~\ref{sec:molecular_benchmarks}) and periodic solids (Sec.~\ref{sec:band_gaps}), comparing our results with those obtained with standard DFT, $G^0W^0$, and other QS$GW$ implementations.
Finally, we summarize our findings and provide an outlook in Sec.~\ref{sec:conclusion}.

\section{\label{sec:Theoretical_Framework}Theoretical Framework}
This section outlines the theoretical foundations of the QS$GW$ method. We begin with a review of the $GW$ approximation and the one-shot $G^0W^0$ approach, followed by a detailed formulation of the QS$GW$ scheme. We then present the NAO-based formalism combined with the LRI technique, which forms the algorithmic basis of our implementation.

\subsection{The \texorpdfstring{$GW$}{GW} approximation and the \texorpdfstring{$G^0W^0$}{G0W0} scheme}
In quantum many-body theory, the central quantity is the single-particle Green's function $G$, which describes the propagation of an electron or hole in an interacting many-body system.

The interacting Green's function $G$ satisfies the Dyson equation, which relates $G$ to the non-interacting Green's function $G^0$ via the self-energy $\Sigma$:
\begin{align}
	 & G(\mathbf{r}, \mathbf{r}', \omega) = \, G^{0}(\mathbf{r}, \mathbf{r}', \omega) \nonumber \\& + \int d\mathbf{r}_1 d\mathbf{r}_2 \, G^{0}(\mathbf{r}, \mathbf{r}_1, \omega) \Sigma(\mathbf{r}_1, \mathbf{r}_2, \omega) G(\mathbf{r}_2, \mathbf{r}', \omega).
	\label{eq:Dysons}
\end{align}
The self-energy $\Sigma$ accounts for all XC effects beyond the mean-field Hartree level and needs to be approximated in practical calculations. The poles of $G$ yield the quasiparticle excitations which, compared to the bare particles (electrons), possess renormalized properties such as modified effective masses and finite lifetimes.

The accuracy of the calculated Green's function depends critically on the approximation chosen for the self-energy.
In this context, the $GW$ approximation (GWA) corresponds to a specific choice of self-energy diagrams, where $\Sigma$ is given by the product of $G$ and the dynamically screened Coulomb interaction $W$.
In the standard one-shot $G^{0}W^{0}$ scheme, the self-energy is evaluated non-self-consistently using an initial $G^0$:
\begin{equation}
	\Sigma(\mathbf{r},\mathbf{r}^{\prime},\omega)=\frac{\mathrm{i}}{2\pi}\int d\omega^{\prime}G^{0}(\mathbf{r},\mathbf{r}^{\prime},\omega+\omega^{\prime})W^{0}(\mathbf{r},\mathbf{r}^{\prime},\omega^{\prime})
	\label{eq:Sigma_G0W0}
\end{equation}
where $W^0$ is the screened Coulomb interaction, typically evaluated within the random phase approximation (RPA).

Concretely, $W^0$ can be expressed as
\begin{equation}
	W^{0}(\mathbf{r},\mathbf{r}^{\prime},\omega)=\int d\mathbf{r}_{1} \varepsilon^{-1}(\mathbf{r},\mathbf{r}_{1},\omega)v(\mathbf{r}_{1},\mathbf{r}^{\prime})\, .
	\label{eq:W0_alt}
\end{equation}
where $\varepsilon^{-1}$ is the inverse of the dielectric function $\varepsilon$, which describes the electron screening process and can be computed as
\begin{equation}
	\varepsilon(\mathbf{r},\mathbf{r}^{\prime},\omega)=\delta(\mathbf{r}-\mathbf{r}^{\prime})-\int d\mathbf{r}_{1}v(\mathbf{r},\mathbf{r}_{1})\chi^{0}(\mathbf{r}_{1},\mathbf{r}^{\prime},\omega) \, .
	\label{eq:dielectric}
\end{equation}
Here, $v(\mathbf{r},\mathbf{r}^{\prime})= 1/|\mathbf{r}-\mathbf{r}^{\prime}|$ is the bare Coulomb interaction and $\chi^0$ is the irreducible polarizability. Under the RPA, $\chi^0$ equals the non-interacting
density response function, given by
\begin{equation}
	\chi^{0}(\mathbf{r},\mathbf{r}^{\prime},\omega)=-\frac{\mathrm{i}}{2\pi}\int d\omega^{\prime}G^{0}(\mathbf{r},\mathbf{r}^{\prime},\omega+\omega^{\prime})G^{0}(\mathbf{r}^{\prime},\mathbf{r},\omega^{\prime}).
	\label{eq:chi0}
\end{equation}
As is evident from the above equations, in the $G^0W^0$ approach, the calculation relies entirely on the non-interacting Green's function $G^0$, which is usually constructed from a preceding DFT calculation:
\begin{equation}
	\begin{aligned}
		G^{0}(\mathbf{r},\mathbf{r}^{\prime},\omega)=\, & \sum_{p}(1-f_{p})\frac{\psi_{p}(\mathbf{r})\psi_{p}^{*}(\mathbf{r}^{\prime})}{\omega+\epsilon_{\rm F}-\epsilon_{p}+ \mathrm{i}\eta} \\
		                                                & +\sum_{p}f_{p}\frac{\psi_{p}(\mathbf{r})\psi_{p}^{*}(\mathbf{r}^{\prime})}{\omega+\epsilon_{\rm F}-\epsilon_{p}- \mathrm{i}\eta}
	\end{aligned}
	\label{eq:G0}
\end{equation}
where $\psi_{p}(\mathbf{r})$ and $\epsilon_{p}$ are the KS orbitals and orbital energies, respectively, $f_p$ is the occupation number, $\epsilon_{\rm F}$ is the Fermi energy (chemical potential), and $\eta$ is a positive infinitesimal.

Within the $G^0W^0$ scheme, the quasiparticle excitation energies are obtained by solving the quasiparticle equation
\begin{equation}
	\epsilon_{p}^{\text{QP}} = \epsilon_{p} + \mathrm{Re} \langle \psi_{p} | \Sigma(\epsilon_{p}^{\text{QP}}) - V^{\mathrm{xc}} | \psi_{p} \rangle,
	\label{eq:QP_G0W0}
\end{equation}
where the self-energy depends on the quasiparticle energy itself.
This nonlinear equation is solved self-consistently for each state, typically by root-finding algorithms.
Importantly, in $G^0W^0$, the Green's function $G^0$ and screened interaction $W^0$ are fixed from the initial DFT calculation and are not updated during the calculation --- hence the term ``one-shot'' $GW$.
Only the diagonal elements of the self-energy matrix are required.

However, as discussed in Sec.~\ref{sec:introduction}, $G^0W^0$ does not provide updated charge densities or quasiparticle wavefunctions beyond the initial KS calculation.
The QS$GW$ method, described in the following subsection, addresses this limitation by providing a self-consistent framework that simultaneously updates both quasiparticle energies and wavefunctions.

\subsection{The \texorpdfstring{QS$GW$}{QSGW} scheme}
The QS$GW$ scheme aims to find an effective static and Hermitian Hamiltonian that best approximates the effective dynamical $GW$ Hamiltonian. In general, we can write such a Hamiltonian as
\begin{equation}
	H^{\mathrm{QS}GW} = -\frac{1}{2}\nabla^2 + V^{\mathrm{ext}}(\mathbf{r}) + V^{\mathrm{H}}(\mathbf{r}) + V^{\mathrm{xc}}\, ,
	\label{eq:H_QSGW}
\end{equation}
where the first three terms -- the kinetic energy operator, the external potential operator, and the Hartree potential operator -- are the same as in the usual KS Hamiltonian, whereas the final term, the XC potential operator $V^{\mathrm{xc}}$, represents a static, Hermitian approximation to
the dynamic $GW$ self-energy $\Sigma^{GW}(\omega)$.
In practice, there are different choices of the actual procedure to construct the XC potential for QS$GW$, with the two most prominent schemes being referred to as ``Mode A'' and ``Mode B'' \cite{VanSchilfgaarde2006}.
Specifically, in ``Mode A'', the XC potential is defined as:

\begin{equation}
	V^{\mathrm{xc}}=\frac{1}{2}\sum_{pp'}\ket{\psi_{p}}\{\mathrm{Re}[\Sigma^{GW}(\epsilon_{p})]_{pp'}+\mathrm{Re}[\Sigma^{GW}(\epsilon_{p'})]_{pp'}\}\bra{\psi_{p'}}\, ,
	\label{eq:Vxc_QSGW}
\end{equation}
whereas in ``Mode B'', it is defined as:
\begin{equation}
	\begin{aligned}
		V^{\mathrm{xc}} = & \sum_{p}\ket{\psi_{p}}\mathrm{Re}[\Sigma^{GW}(\epsilon_{p})]_{pp}\bra{\psi_{p}}             \\
		                  & + \sum_{p,p\neq p'}\ket{\psi_{p}}\mathrm{Re}[\Sigma^{GW}(\epsilon_{\rm F})]_{pp'}\bra{\psi_{p'}}.
	\end{aligned}
	\label{eq:Vxc_QSGW_B}
\end{equation}
Here $\epsilon_{p}$ is the $p$-th eigenvalue of the Hamiltonian $H^{\mathrm{QS}GW}$, $\epsilon_{\rm F}$ is the Fermi energy, and ``$\mathrm{Re}$'' denotes the Hermitian part of the self-energy matrix, i.e., $\mathrm{Re}[A]_{pp'} \equiv (A_{pp'} + A_{p'p}^{*})/2$. 
Formally, the construction in ``Mode A'' can also be derived by minimizing the length of the gradient of the
Klein total energy functional over non-interacting Green's functions (given by $(\omega -H^{\mathrm{QS}GW} )^{-1}$), as demonstrated by Ismail-Beigi \cite{Ismail-Beigi2017}.

In a practical QS$GW$ calculation, the following steps need to be executed:

\begin{enumerate}
	\item Start with an initial guess for the effective potential $V^{\mathrm{xc}}$, usually taken from a preceding DFT calculation.
	\item Construct the effective Hamiltonian $H^{\mathrm{QS}GW}$ using the current $V^{\mathrm{xc}}$ and solve the eigenvalue problem to obtain the QS$GW$ orbitals $\psi_{p}(\mathbf{r})$ and eigenvalues $\epsilon_{p}$. 
	\item Build the non-interacting Green's function $G^0$ from the QS$GW$ orbitals and eigenvalues.
	\item Calculate the self-energy $\Sigma(\mathbf{r},\mathbf{r}^{\prime},\omega)$ using the $GW$ approximation based on $G^0$ as described in Eqs.~(\ref{eq:Sigma_G0W0})--(\ref{eq:G0}).
	\item Update the effective potential $V^{\mathrm{xc}}$ using the chosen QS$GW$ mode (A or B) to obtain a new static and Hermitian approximation of the self-energy.
	\item Check for convergence. If the change in the quasiparticle energies is below a predefined threshold, the calculation is considered converged. Otherwise, return to step 2.
\end{enumerate}

The above-outlined QS$GW$ scheme has been implemented in the LibRPA package \cite{shi2025librpa,Zhang2026arXiv}.
LibRPA was originally developed for the RPA correlation energy \cite{shi2025librpa} and one-shot $G^0W^0$ calculations \cite{Zhang2026arXiv}.
LibRPA is interfaced with FHI-aims to acquire necessary inputs from DFT calculations. We use the NAO-based real-space formalism combined with the LRI technique to construct the XC potential operator of QS$GW$ efficiently, which will be detailed in the next subsection.

\subsection{NAO-based formalism for \texorpdfstring{QS\textit{GW}}{QSGW}}

To implement the QS$GW$ formalism efficiently within the NAO framework, we employ a space-time approach \cite{Ren2021PRM,Zhang2026arXiv} combined with the LRI technique \cite{lin2020accuracy}.
This approach facilitates the computation of the full $GW$ self-energy matrix required by the QS$GW$ scheme. Furthermore, the use of NAO basis functions and the LRI technique gives rise to significant sparsity in the LRI expansion coefficients. Exploiting this sparsity leads to an $\mathcal{O}(N^2)$-scaling algorithm for constructing the $GW$ self-energy, compared to the $\mathcal{O}(N^4)$ scaling in the conventional approach.
A detailed description of the underlying $G^0W^0$ implementation has been provided in a separate work \cite{Zhang2026arXiv}. Here, we only summarize the key steps and equations specific to the QS$GW$ implementation.

\subsubsection{NAO basis set representation}

First, the single-particle orbitals (KS orbitals in the initial iteration, or the QS$GW$ orbitals in subsequent iterations) are expanded in a basis of numerical atomic orbitals (NAOs):
\begin{equation}
	\psi_{p\mathbf{k}}(\mathbf{r}) = \sum_i c_{p\mathbf{k}}^i \varphi_{i\mathbf{k}}(\mathbf{r})
\end{equation}
Here, $c_{p\mathbf{k}}^i$ are the expansion coefficients for the $p$-th state at Bloch wavevector $\mathbf{k}$ in terms of the NAO basis. The Bloch sums of NAOs are defined as:
\begin{equation}
	\varphi_{i\mathbf{k}}(\mathbf{r}) = \frac{1}{\sqrt{N_{\mathbf{k}}}} \sum_{\mathbf{R}} e^{\mathrm{i}\mathbf{k}\cdot\mathbf{R}} \varphi_{i\mathbf{R}}(\mathbf{r}),
\end{equation}
with each localized NAO taking the form:
\begin{equation}
	\varphi_{i\mathbf{R}}(\mathbf{r}) = u_{\kappa l}(|\mathbf{r} - \mathbf{t}_I - \mathbf{R}|) S_{lm}(\widehat{\mathbf{r} - \mathbf{t}_I - \mathbf{R}}).
\end{equation}
Here, $u_{\kappa l}$ denotes the radial part of the NAO and $S_{lm}$ corresponds to the real spherical harmonics. The composite index $i = \{I, \kappa, l, m\}$ specifies the atomic center $I$, the radial function index $\kappa$, and the angular quantum numbers $(l, m)$.
Furthermore, $\mathbf{R}$ is the lattice translation vector, and $\mathbf{t}_I$ represents the position of atom $I$ within the unit cell.

\subsubsection{Localized resolution of identity technique}

For QS$GW$, we employ an auxiliary basis set to represent the two-electron quantities and LRI to decompose the four-index tensors into three- and two-index ones, similar to the canonical $G^0W^0$ implementation for periodic systems in FHI-aims \cite{Ren2021PRM} and our low-scaling RPA and $G^0W^0$ implementation in LibRPA \cite{shi2024subquadratic,shi2025librpa,Zhang2026arXiv}.
In this approach, a key step is to expand the products of NAOs in terms of auxiliary basis functions (ABFs):
\begin{align}
	 & \varphi_{i\mathbf{R}}(\mathbf{r}) \varphi_{j\mathbf{R}'}(\mathbf{r}) \approx \sum_{\mu \in I} C_{i(\mathbf{R}), j(\mathbf{R}')}^{\mu(\mathbf{R})} P_{\mu\mathbf{R}}(\mathbf{r}) \nonumber \\
	 & + \sum_{\mu \in J} C_{i(\mathbf{R}), j(\mathbf{R}')}^{\mu(\mathbf{R}')} P_{\mu\mathbf{R}'}(\mathbf{r}),
	\label{eq:LRI_expansion}
\end{align}
where $P_{\mu\mathbf{R}}(\mathbf{r})$ denotes the ABFs which are required to be centered on atoms $I$ or $J$, on which the two NAOs are located. In Eq.~\ref{eq:LRI_expansion}, $C_{i(\mathbf{R}), j(\mathbf{R}')}^{\mu(\mathbf{R})}$ are the LRI expansion coefficients, and the lattice vector $\mathbf{R}$ in  parentheses specifies the unit cell where the NAO $i$ or ABF $\mu$ is localized in a periodic system.

Now, by substituting this expansion into the four-orbital electron repulsion integrals, these computationally expensive integrals are decomposed into three-index tensors (involving two NAOs and one ABF) and two-index matrices (involving two ABFs). This decomposition constitutes the foundation for the efficient evaluation of self-energy terms, as explicitly demonstrated later in the construction of the correlation self-energy [see Eq.~(\ref{eq:sigmac-R})].

\subsubsection{Space-time formalism of \texorpdfstring{QS$GW$}{QSGW} with NAO and LRI}

The non-interacting Green's function in real space and imaginary time is expressed in the NAO basis as:
\begin{equation}
	G^0(\mathbf{r}, \mathbf{r}', \mathrm{i}\tau) = \sum_{ij} \sum_{\mathbf{R}, \mathbf{R}'} \varphi_{i\mathbf{R}}(\mathbf{r}) G_{ij}^0(\mathbf{R}' - \mathbf{R}, \mathrm{i}\tau) \varphi_{j\mathbf{R}'}(\mathbf{r}'),
\end{equation}
with the matrix elements given by

\begin{equation}
	G_{ij}^0(\mathbf{R}, \mathrm{i}\tau) = \frac{1}{N_{\mathbf{k}}} \sum_{p\mathbf{k}} e^{-\mathrm{i}\mathbf{k}\cdot\mathbf{R}} c_{p\mathbf{k}}^i c_{p\mathbf{k}}^{j*} \Xi_{p\mathbf{k}}(\tau),
\end{equation}
where
\begin{equation}
	\Xi_{p\mathbf{k}}(\tau) = e^{-(\epsilon_{p\mathbf{k}} - \epsilon_{\rm F})\tau}
	\left[\theta(\tau)(1 - f_{p\mathbf{k}}) - \theta(-\tau)f_{p\mathbf{k}} \right].
\end{equation}
Here $\theta(\tau)$ is the Heaviside step function, $f_{p\mathbf{k}}$ is the occupation number, and $\epsilon_{\rm F}$ is the Fermi energy (chemical potential).  For the sake of simplicity, only the spin-degenerate case with integer orbital occupations is considered. The extension to
the spin polarized case is straightforward.

Using LRI, the irreducible polarizability in real space and imaginary time is expanded in terms of ABFs,
\begin{equation}
	\chi^0(\mathbf{r}, \mathbf{r}', \mathrm{i}\tau) = \sum_{\mu\nu} \sum_{\mathbf{R}, \mathbf{R}'} P_{\mu\mathbf{R}}(\mathbf{r}) \chi_{\mu\nu}^0(\mathbf{R}' - \mathbf{R}, \mathrm{i}\tau) P_{\nu\mathbf{R}'}(\mathbf{r}'),
\end{equation}
where the polarizability matrix is computed as (see Refs.~\cite{shi2024subquadratic,shi2025librpa}):

\begin{align}
  \chi^{0}_{\mu\nu}(\mathbf{R},\mathrm{i}\tau)
  &= -\sum_{ij}\sum_{kl}\sum_{\mathbf{R}_1,\mathbf{R}_2}
  C^{\mu(\mathbf{0})}_{i(\mathbf{0}),k(\mathbf{R}_1)}\,
  C^{\nu(\mathbf{R})}_{j(\mathbf{R}),l(\mathbf{R}_2)}
  \nonumber\\
  &\quad\times\Big[
  G^{0}_{ij}(\mathbf{R},\mathrm{i}\tau)\,
  G^{0}_{lk}(\mathbf{R}_1-\mathbf{R}_2,-\mathrm{i}\tau)
  \nonumber\\
  &\qquad
  +G^{0}_{il}(\mathbf{R}_2,\mathrm{i}\tau)\,
  G^{0}_{jk}(\mathbf{R}_1-\mathbf{R},-\mathrm{i}\tau)
  \nonumber\\
  &\qquad
  +G^{0}_{ji}(-\mathbf{R},-\mathrm{i}\tau)\,
  G^{0}_{kl}(\mathbf{R}_2-\mathbf{R}_1,\mathrm{i}\tau)
  \nonumber\\
  &\qquad
  +G^{0}_{li}(-\mathbf{R}_2,-\mathrm{i}\tau)\,
  G^{0}_{kj}(\mathbf{R}-\mathbf{R}_1,\mathrm{i}\tau)
  \Big].
  \label{eq:chi0-mn}
  \end{align}
Here, $C_{i(\mathbf{0}), k(\mathbf{R}')}^{\mu(\mathbf{0})}$ are the LRI expansion coefficients introduced in Eq.~\ref{eq:LRI_expansion}. In practical calculations, $C_{i(\mathbf{0}), k(\mathbf{R}')}^{\mu(\mathbf{0})}$ are provided by FHI-aims, interfaced with LibRPA.

The next key quantity is the screened Coulomb matrix represented within the ABF basis, which is
computed as:
\begin{equation}
	W_{\mu\nu}^0(\mathbf{q}, \mathrm{i}\omega) = \sum_{\mu'\nu'} V_{\mu\mu'}^{1/2}
	\left[\tilde{\varepsilon}^{-1}(\mathbf{q}, \mathrm{i}\omega)\right]_{\mu'\nu'}
	V_{\nu'\nu}^{1/2}\, ,
\end{equation}
where $V^{1/2}$ is the square root of the bare Coulomb matrix $V$, and ${\varepsilon}^{-1}(\mathbf{q}, \mathrm{i}\omega)$ is the inverse of the symmetrized dielectric
function $\varepsilon$, given by

\begin{equation}
	\tilde{\varepsilon}_{\mu\nu}(\mathbf{q}, \mathrm{i}\omega) = \delta_{\mu\nu} - \sum_{\mu'\nu'} V_{\mu\mu'}^{1/2}(\mathbf{q}) \chi_{\mu'\nu'}^0(\mathbf{q}, \mathrm{i}\omega) V_{\nu'\nu}^{1/2}(\mathbf{q}).
	\label{eq:dielectric_function}
\end{equation}
Here $\mathbf{q}$ denotes the momentum transfer (wave vector of the dielectric response), equivalent to
${\mathbf{k - k'}}$. For a uniform ${\mathbf{k}}$ mesh, the ${\mathbf{q}}$ mesh is the same as the ${\mathbf{k}}$ one, but they can differ in general cases.

To calculate the dielectric function using Eq.~\ref{eq:dielectric_function}, the irreducible polarization
function $\chi^0$ needs to be Fourier transformed from the imaginary time axis to the imaginary frequency axis.
This can be efficiently done using MiniMax time/frequency grids \cite{kaltak2014minimax,azizi2023greenx}.

In the $GW$ approximation, the self-energy can be decomposed into a time-independent  exchange part $\Sigma^{\rm x}$ and a time-dependent correlation part $\Sigma^{\rm c}$, i.e., $\Sigma=\Sigma^{\rm x}+\Sigma^{\rm c}$. 
These two parts correspond to splitting the screened Coulomb interaction $W$ into the bare Coulomb interaction $V$ and a remaining dynamic part $W^c$, i.e., $W=V+W^c$. Accordingly, we can compute these two parts within the NAO basis as follows:
\begin{equation}
	\begin{aligned}
		\Sigma_{pp'}^{\mathrm{x}}(\mathbf{k})                & = \sum_{ij}c_{p\mathbf{k}}^{i*}c_{p'\mathbf{k}}^{j}\sum_{\mathbf{R}}\mathrm{e}^{\mathrm{i}\mathbf{k}\cdot\mathbf{R}}\Sigma_{ij}^{\mathrm{x}}(\mathbf{R}),                \\
		\Sigma_{pp'}^{\mathrm{c}}(\mathbf{k},\mathrm{i}\tau) & = \sum_{ij}c_{p\mathbf{k}}^{i*}c_{p'\mathbf{k}}^{j}\sum_{\mathbf{R}}\mathrm{e}^{\mathrm{i}\mathbf{k}\cdot\mathbf{R}}\Sigma_{ij}^{\mathrm{c}}(\mathbf{R},\mathrm{i}\tau).
	\end{aligned}
	\label{eq:Sigma_x}
\end{equation}
Efficient algorithms to calculate the exact-exchange matrix in real space, $\Sigma_{ij}^{\mathrm{x}}(\mathbf{R})$, have been extensively discussed previously \cite{lin2020accuracy,Lin/Ren/He:2021,lin2025efficient}, and will not be further discussed here.
As for the correlation part, the matrix elements $\Sigma^{\rm c}_{ij}(\mathbf{R},\mathrm{i}\tau)$ can be expressed in terms of the LRI coefficients, the screened Coulomb interaction $W_{\mu\nu}^c(\mathbf{R},\mathrm{i}\tau)$, and the Green's function as:
\begin{equation}
	\begin{aligned}
		 & \Sigma^{\rm c}_{ij}(\mathbf{R},\mathrm{i}\tau) = -\sum_{k\mathbf{R}_1}\sum_{l\mathbf{R}_2} G_{kl}^{0}(\mathbf{R}_2-\mathbf{R}_1,\mathrm{i}\tau)                                                                         \\
		 & \times \left[\sum_{\mu\in I}\sum_{\nu\in J} C^{\mu(\mathbf{0})}_{i(\mathbf{0}),k(\mathbf{R}_1)} W_{\mu\nu}^{c}(\mathbf{R},\mathrm{i}\tau) C^{\nu(\mathbf{R})}_{j(\mathbf{R}),l(\mathbf{R}_2)} \right.                   \\
		 & + \,\sum_{\mu\in K}\sum_{\nu\in J} C^{\mu(\mathbf{R}_1)}_{i(\mathbf{0}),k(\mathbf{R}_1)} W_{\mu\nu}^{c}(\mathbf{R}-\mathbf{R}_1,\mathrm{i}\tau) C^{\nu(\mathbf{R})}_{j(\mathbf{R}),l(\mathbf{R}_2)}                     \\
		 & + \,\sum_{\mu\in I}\sum_{\nu\in L} C^{\mu(\mathbf{0})}_{i(\mathbf{0}),k(\mathbf{R}_1)} W_{\mu\nu}^{c}(\mathbf{R}_2,\mathrm{i}\tau) C^{\nu(\mathbf{R}_2)}_{j(\mathbf{R}),l(\mathbf{R}_2)}                                \\
		 & + \left. \,\sum_{\mu\in K}\sum_{\nu\in L} C^{\mu(\mathbf{R}_1)}_{i(\mathbf{0}),k(\mathbf{R}_1)} W_{\mu\nu}^{c}(\mathbf{R}_2-\mathbf{R}_1,\mathrm{i}\tau) C^{\nu(\mathbf{R}_2)}_{j(\mathbf{R}),l(\mathbf{R}_2)} \right].
	\end{aligned}
	\label{eq:sigmac-R}
\end{equation}

The screened Coulomb matrix in real space and imaginary time $W_{\mu\nu}^c(\mathbf{R}, \mathrm{i}\tau)$ is obtained from its counterpart in
$\mathbf{q}$ space and imaginary frequency via Fourier transform. Further details of our implementation
are presented in Ref.~\cite{Zhang2026arXiv}.

Thus, we have constructed all the necessary components to compute the dynamic self-energy $\Sigma(\omega)$ within the NAO and LRI framework using the space-time formalism.
To obtain the self-energy on the real frequency axis, we perform an analytic continuation from imaginary frequencies using the Pad\'e approximant method \cite{vidberg1977solving}.

The final step in our QS$GW$ implementation is the construction of the static, Hermitian XC potential matrix $V^{\mathrm{xc}}$ that approximates the dynamical self-energy $\Sigma(\omega)$. The self-energy matrix elements in the eigenbasis are computed as:

\begin{equation}
	\Sigma_{pp'}(\mathbf{k},\omega) = \langle \psi_{p\mathbf{k}} | \Sigma(\omega) | \psi_{p'\mathbf{k}} \rangle = \sum_{ij} c_{p\mathbf{k}}^{i*} c_{p'\mathbf{k}}^j \Sigma_{ij}(\mathbf{k},\omega),
\end{equation}
where $\Sigma_{ij}(\mathbf{k},\omega)$ are the matrix elements in the NAO basis and  $\Sigma_{pp'}(\mathbf{k},\omega)$ are those in the eigenbasis of the QS$GW$ Hamiltonian.

Subsequently, the QS$GW$ XC potential $V^{\mathrm{xc}}$ is constructed according to either ``Mode A'' or ``Mode B'', as defined in Eqs.~(\ref{eq:Vxc_QSGW}) and (\ref{eq:Vxc_QSGW_B}).

The combination of NAOs, LRI, and the space-time formalism provides an efficient framework for QS$GW$ calculations, maintaining accuracy while keeping the computational cost tractable for large extended systems. 

\section{\label{sec:implementation}Implementation details}

\subsection{\label{QSGW_in_LibRPA} \texorpdfstring{QS\textit{GW}}{QSGW} in LibRPA}
\begin{figure}[htbp]

	\includegraphics[width=1\linewidth]{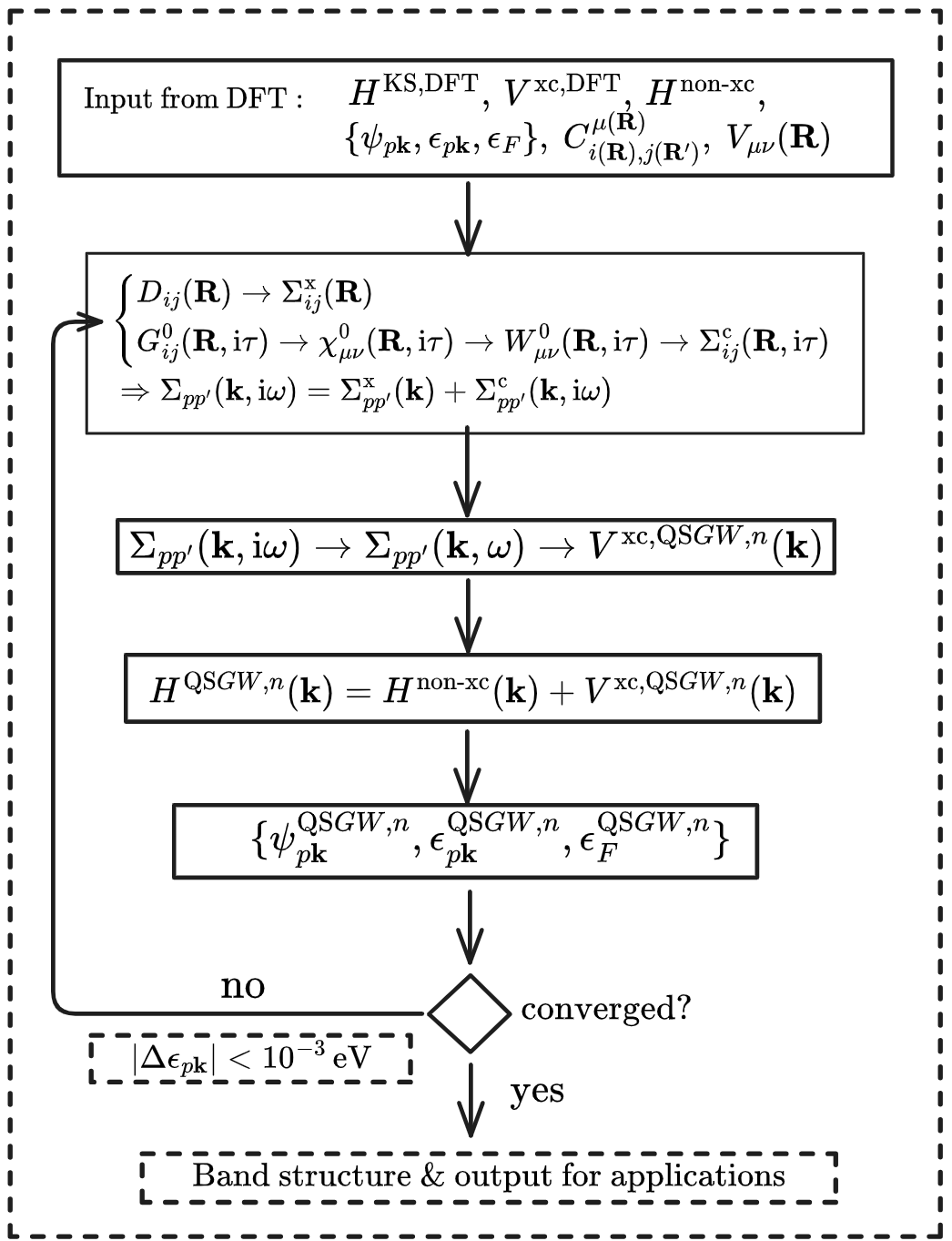}

	\caption {Workflow of the QS$GW$ calculation, starting from a KS-DFT calculation and converging to self-consistency using the LibRPA software. Indices $i,j$ label NAO basis functions (composite index $i=\{I,\kappa,l,m\}$ with magnetic quantum number $m$ in $S_{lm}$); $p,p'$ label single-particle eigenstates $\psi_{p\mathbf{k}}$; $\mathbf{k}$ is the Bloch wave vector; $\mathbf{q}$ is the momentum transfer used in $W_{\mu\nu}^0(\mathbf{q},\mathrm{i}\omega)$; $I,J$ represent atom indices; and $n$ is the iteration number.} 
	\label{fig:QSGW_workflow}
\end{figure}

Our QS$GW$ implementation is built upon the existing space-time $G^{0}W^{0}$ framework within the LibRPA software package \cite{Zhang2026arXiv}.
Here, we focus on the specific extensions required for the self-consistent scheme, as illustrated in Fig.~\ref{fig:QSGW_workflow}.

First, a standard KS-DFT calculation is performed to obtain an initial Kohn-Sham Hamiltonian $H^{\text{KS,DFT}}$, along with the corresponding orbitals and eigenvalues. We employ FHI-aims as the DFT engine to generate the initial input for LibRPA. At the beginning of the QS$GW$ iteration, the non-XC part of the Hamiltonian is extracted from the KS-DFT calculation and remains fixed throughout the QS$GW$ iterations. Namely, the DFT XC potential is first subtracted from the KS Hamiltonian:
\begin{equation}
	H^{\text{non-xc}} =  -\frac{1}{2}\nabla^2 + V^{\mathrm{ext}}(\mathbf{r}) + V^{\mathrm{H}}(\mathbf{r}) = H^{\text{KS,DFT}} - V^{\mathrm{xc,DFT}},
	\label{eq:H_non_xc}
\end{equation}
and the resultant $H^{\text{non-xc}}$ will be combined with the $V^{\mathrm{xc}}$ matrix derived from the
$GW$ self-energy to build the QS$GW$ Hamiltonian.

Subsequently, the QS$GW$ iteration proceeds analogously to a standard $G^0W^0$ calculation in LibRPA until the self-energy on the imaginary frequency axis, $\Sigma_{pp'}(\mathbf{k}, \mathrm{i}\omega)$, is obtained.

A Pad\'e approximation is then employed to perform the analytic continuation from the imaginary frequency axis to the real frequency axis, yielding the self-energy at real frequencies, $\Sigma_{pp'}(\mathbf{k}, \omega)$. A detailed discussion on the stability of this analytic continuation is provided in Sec.~\ref{analy}. 

Once the dynamic self-energy is obtained, the static and Hermitian XC potential $V^{\mathrm{xc}}$ is constructed according to either ``Mode A'' or ``Mode B'', as defined in Eqs.~(\ref{eq:Vxc_QSGW}) and (\ref{eq:Vxc_QSGW_B}).

Finally, the updated orbitals and eigenvalues are fed back into LibRPA to execute the next QS$GW$ iteration. This process is repeated until the quasiparticle energies are converged within a predefined threshold, typically set to $10^{-3}$~eV in our calculations. The $n$-th QS$GW$ Hamiltonian can thus be expressed as:

\begin{equation}
	H^{\mathrm{QS}GW,n}(\mathbf{k}) = H^{\text{non-xc}}(\mathbf{k}) + V^{\mathrm{xc,QS}GW,n}(\mathbf{k}).
	\label{eq:H_QSGW_n}
\end{equation}
To improve the stability of the self-consistency cycle for molecular systems, we employ Pulay's DIIS mixing method \cite{Pulay1980} when updating the QS$GW$ Hamiltonian. For periodic systems, the iterations are generally stable in our benchmarks and converge without DIIS mixing. Our QS$GW$ implementation typically requires about 10 iterations to achieve convergence of eigenvalues within $10^{-3}$~eV for periodic systems, whereas molecular systems may still require tens to hundreds of iterations. Upon convergence, the final quasiparticle energies and orbitals are utilized to compute various electronic properties, such as band structures and topological invariants.

\subsection{\label{analy} Analytic continuation of self-energy}
In this work, we employ the Pad\'e approximation to perform the analytic continuation of the self-energy from the imaginary frequency axis to the real frequency axis.
As previously reported \cite{Forster2021}, the Pad\'e analytic continuation in QS$GW$ is significantly more sensitive to numerical noise compared to standard $G^0W^0$ calculations, which can affect the stability and accuracy of the results.
This sensitivity arises from the fact that standard $G^0W^0$ calculations require only the diagonal elements of the self-energy matrix for quasiparticle energy calculations, whereas in QS$GW$, the full self-energy matrix is
needed to construct the static XC potential $V^{\mathrm{xc}}$.
The off-diagonal elements of the self-energy matrix are usually small in magnitude, making them particularly susceptible to numerical noise introduced during the analytic continuation process.

Theoretically, in the absence of numerical noise, the analytic continuation of $\Sigma(\mathrm{i}\omega)$ should yield identical results regardless of the basis representation (e.g., the initial KS basis or the updated QS$GW$ eigenbasis).
In this section, ``basis'' refers to this representation choice for $\Sigma$ and its analytic continuation, and should not be confused with the underlying NAO basis settings (e.g., \texttt{intermediate\_gw}, \texttt{really\_tight}).
However, numerical noise inevitably introduces a basis representation dependence.
To evaluate the impact of this dependence, particularly regarding different QS$GW$ construction schemes, we performed calculations on Si and MgO using both ``Mode A'' and ``Mode B''. We compared the results obtained in two different basis representations : the initial KS basis from DFT and the updated QS$GW$ eigenbasis (obtained by diagonalizing the QS$GW$ Hamiltonian from the previous iteration, as defined in Eq.~(\ref{eq:H_QSGW_n})).

\begin{figure}[!htbp]
	\centering
	\includegraphics[width=1\linewidth]{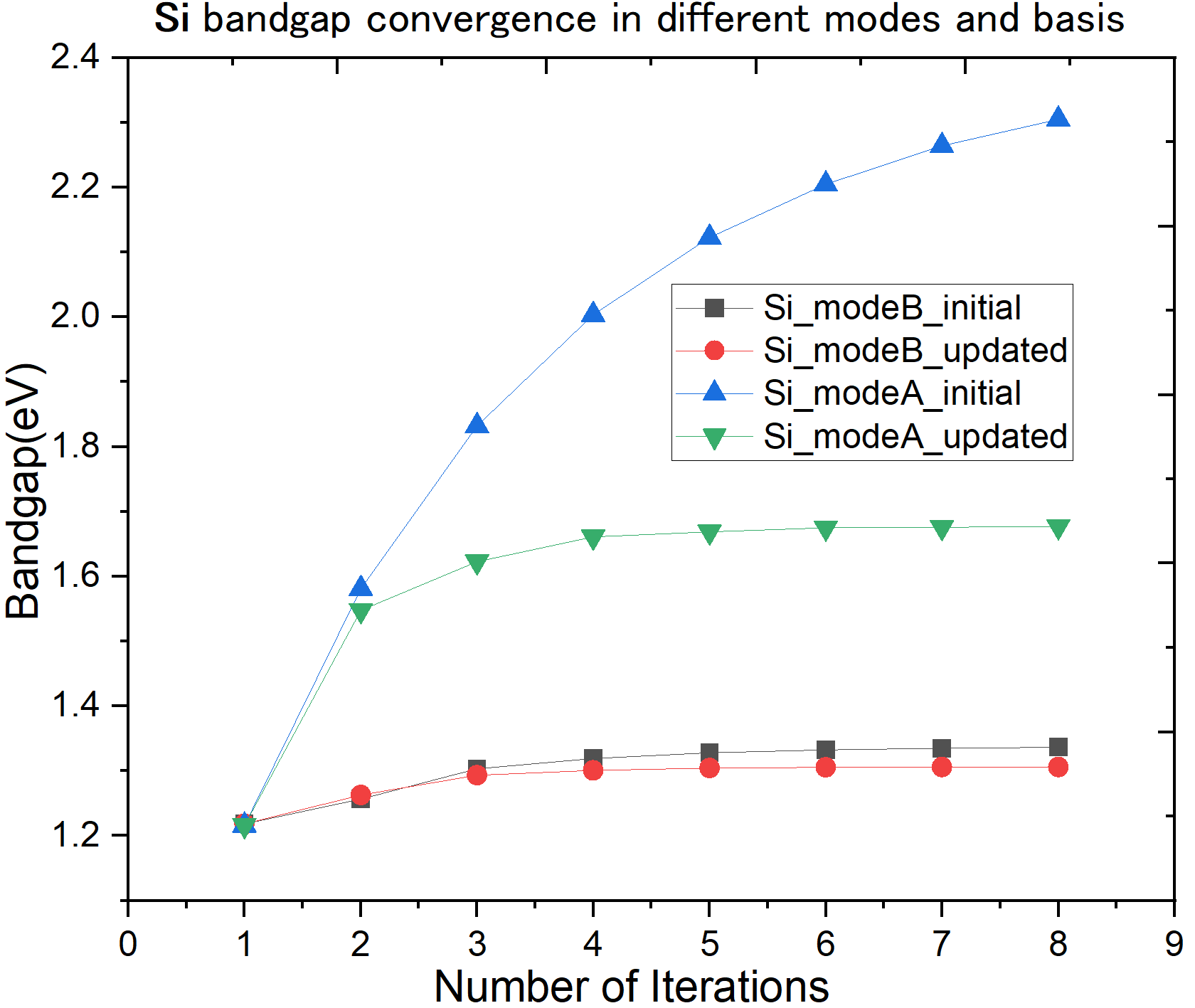}
	\includegraphics[width=1\linewidth]{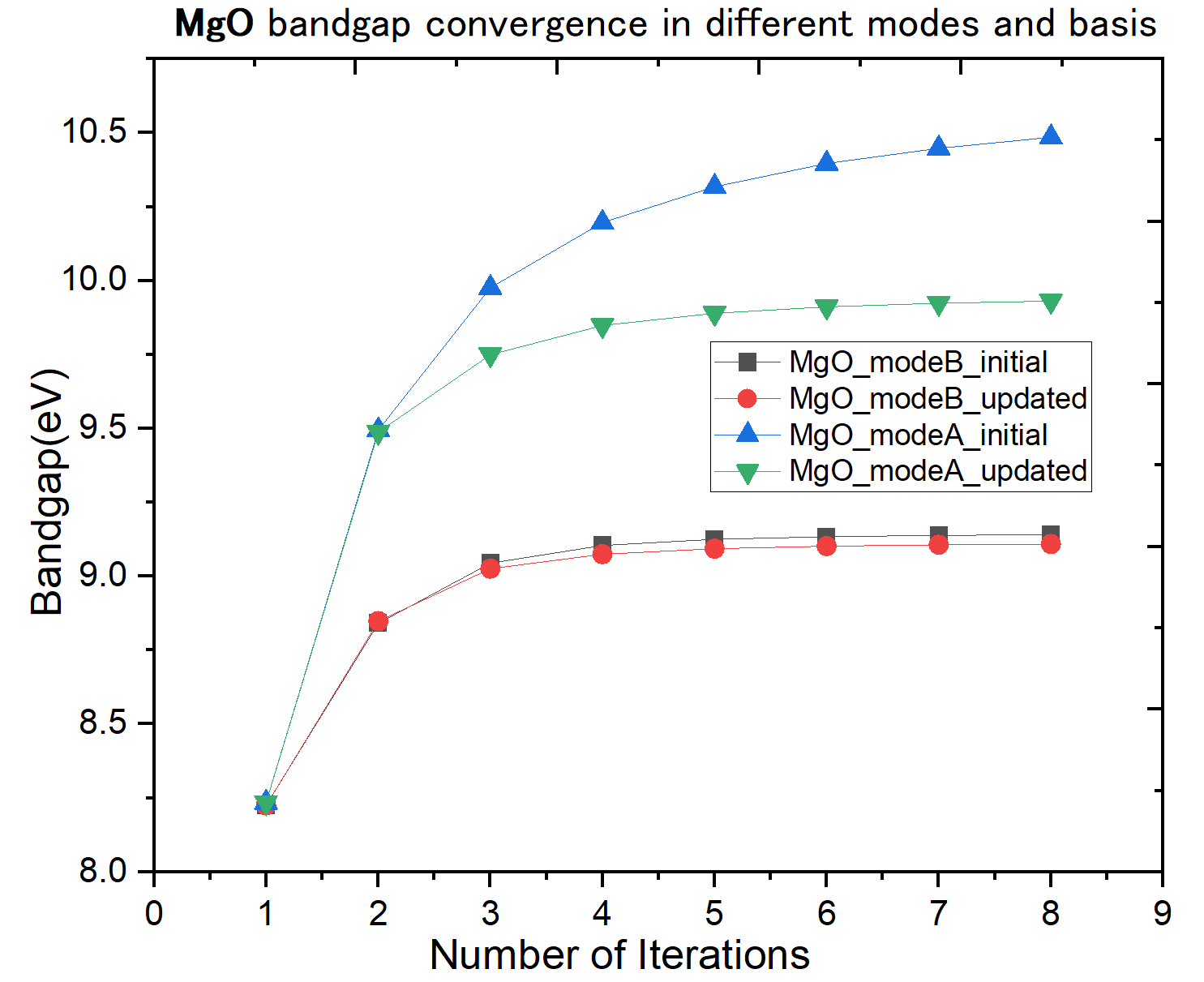}
	\caption{Basis representation dependence tests of analytic continuation in QS$GW$ for ``Mode A'' and ``Mode B'' for Si (top) and MgO (bottom). The Si calculations employed a $6\times6\times6$ k-point mesh, while MgO employed an $8\times8\times8$ mesh. Both used a frequency grid of 16 points and the \texttt{intermediate\_gw} basis sets. In the legend, ``initial'' denotes analytic continuation performed in the initial KS-DFT eigenbasis, while ``updated'' denotes analytic continuation performed in the updated QS$GW$ eigenbasis from the previous iteration. }
	\label{fig:si_mgo_basis_analy}
\end{figure}

As illustrated in Fig.~\ref{fig:si_mgo_basis_analy}, ``Mode A'' exhibits significant basis representation dependence for both Si and MgO. Notably, when the iteration number exceeds 3, the band gaps calculated using ``Mode A'' in different basis representations begin to diverge significantly.
In contrast, ``Mode B'' demonstrates much better consistency between different basis representations, implying that the analytic continuation in ``Mode B'' is far less sensitive to numerical noise.
Specifically, after 5 iterations, the difference in band gaps calculated using the two different basis representations for ``Mode B'' converges to within $10^{-2}$~eV for both materials.
Furthermore, ``Mode B'' exhibits faster convergence rates compared to ``Mode A''. Consequently, while ``Mode A'' may require more robust analytic continuation techniques---such as the maximum entropy method or stochastic optimization---to mitigate noise, ``Mode B'' proves to be numerically stable within our current framework.

Based on these tests for Si and MgO, we employ ``Mode B'' and perform analytic continuation in the initial KS basis for all benchmark calculations reported below.

However, we note that even ``Mode B'' exhibits a small but non-negligible basis representation dependence, as evidenced in Fig.~\ref{fig:si_mgo_basis_analy}. For Si, the band gaps converged in the KS basis and the updated QS$GW$ eigenbasis differ by approximately 0.03 eV after 8 iterations (1.337 eV vs. 1.306 eV). A similar behavior is observed for MgO, with a difference of about 0.03 eV between the two basis representations (9.140 eV vs. 9.109 eV). We treat this residual spread as an estimate of the numerical uncertainty associated with the analytic-continuation protocol used here.

\subsection{\label{sec:convergence} Convergence Behavior}
To ensure the reliability and consistency of the QS$GW$ results, we now proceed to discuss several key numerical parameters: the ${\bf k}$-point mesh, frequency grid, and basis settings.

\subsubsection{\label{sec:kpoint_freq_convergence} \texorpdfstring{$\mathbf{k}$}{k}-point and frequency grid convergence}

\begin{figure}[!htbp]
	\centering
	\includegraphics[width=\linewidth,height=0.45\textheight,keepaspectratio]{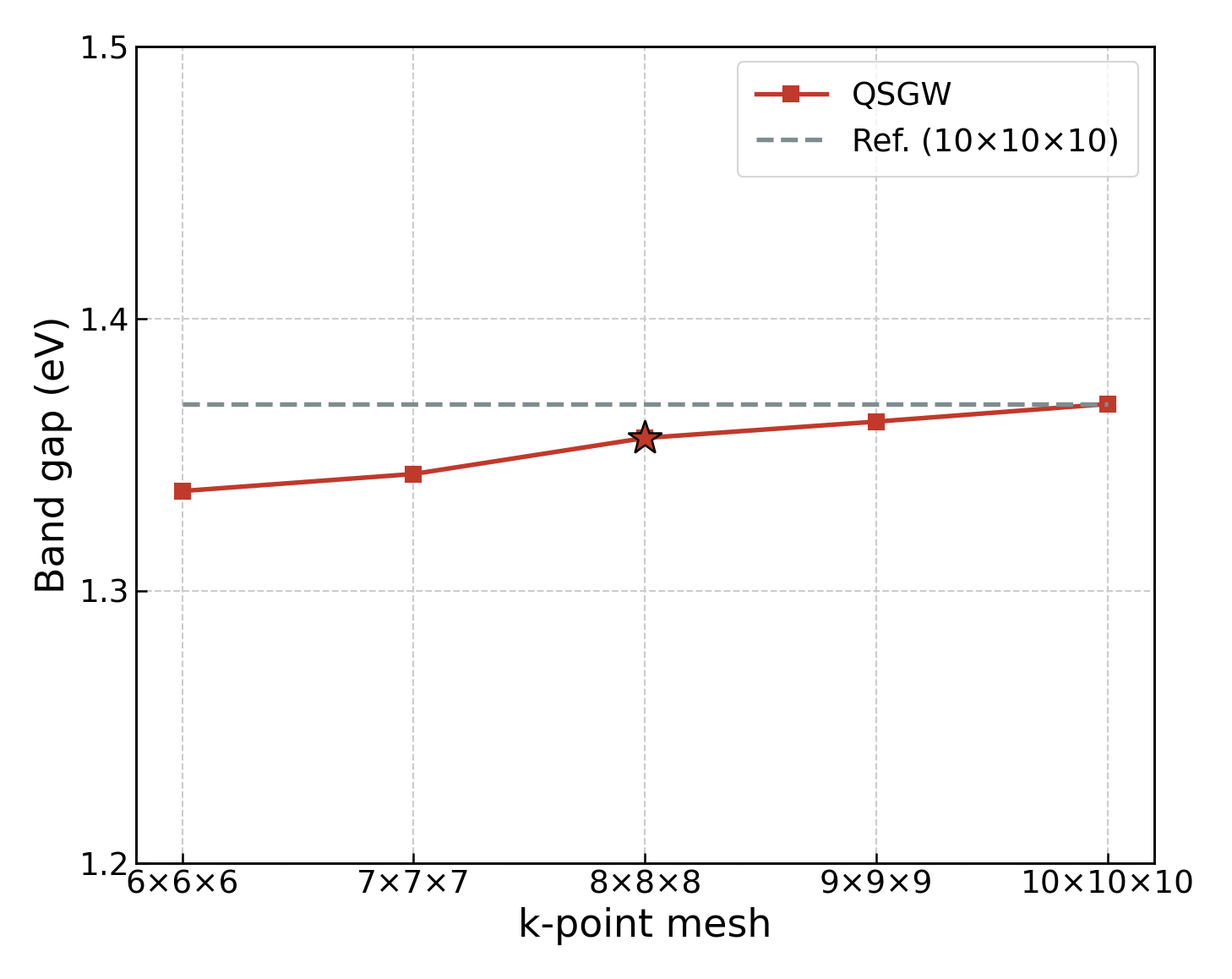}
	\caption{k-point convergence of the Si QS$GW$ band gap with respect to uniform $n\times n\times n$ ${\bf k}$-point meshes ($n=6$--10) using the \texttt{intermediate\_gw} basis sets. The $10\times10\times10$ mesh is taken as the reference.}
	\label{fig:si_kpoint_convergence}
\end{figure}

First, taking Si as an example, we check the convergence of the QS$GW$ band gap with respect to the Brillouin-zone sampling using uniform $n\times n\times n$ ${\bf k}$-point meshes ($n=6$--10), and the results are shown in Fig.~\ref{fig:si_kpoint_convergence}. Taking the $10\times10\times10$ mesh as the reference, the residual QS$GW$ band-gap difference decreases from 32~meV at $6\times6\times6$ to 6.5~meV at $9\times9\times9$ (with $E_g^{\mathrm{QS}GW}=1.3621$~eV at $9\times9\times9$ and $E_g^{\mathrm{QS}GW}=1.3686$~eV at $10\times10\times10$). At $8\times8\times8$ ${\bf k}$-mesh, the difference from the reference value at $10\times10\times10$ is 12.5~meV. This level of convergence is not fully satisfactory, but it represents 
a good compromise between numerical accuracy and computational load; hence $8\times8\times8$ ${\bf k}$-mesh is adopted in this work for cubic systems in production calculations.

In the present implementation, a likely reason for the relatively slow convergence with respect to ${\bf k}$-mesh is that the head correction term (i.e., the $\Gamma$-point correction to the dielectric matrix; see Ref.~\cite{Zhang2026arXiv}) is updated only in the first QS$GW$ iteration and is kept fixed in subsequent iterations.  We are still working on this issue by updating the treatment of the head correction term at each iteration and expect that this will improve the behavior of the ${\bf k}$-point convergence.

\begin{figure}[!htbp]
	\centering
	\includegraphics[width=\linewidth,height=0.45\textheight,keepaspectratio]{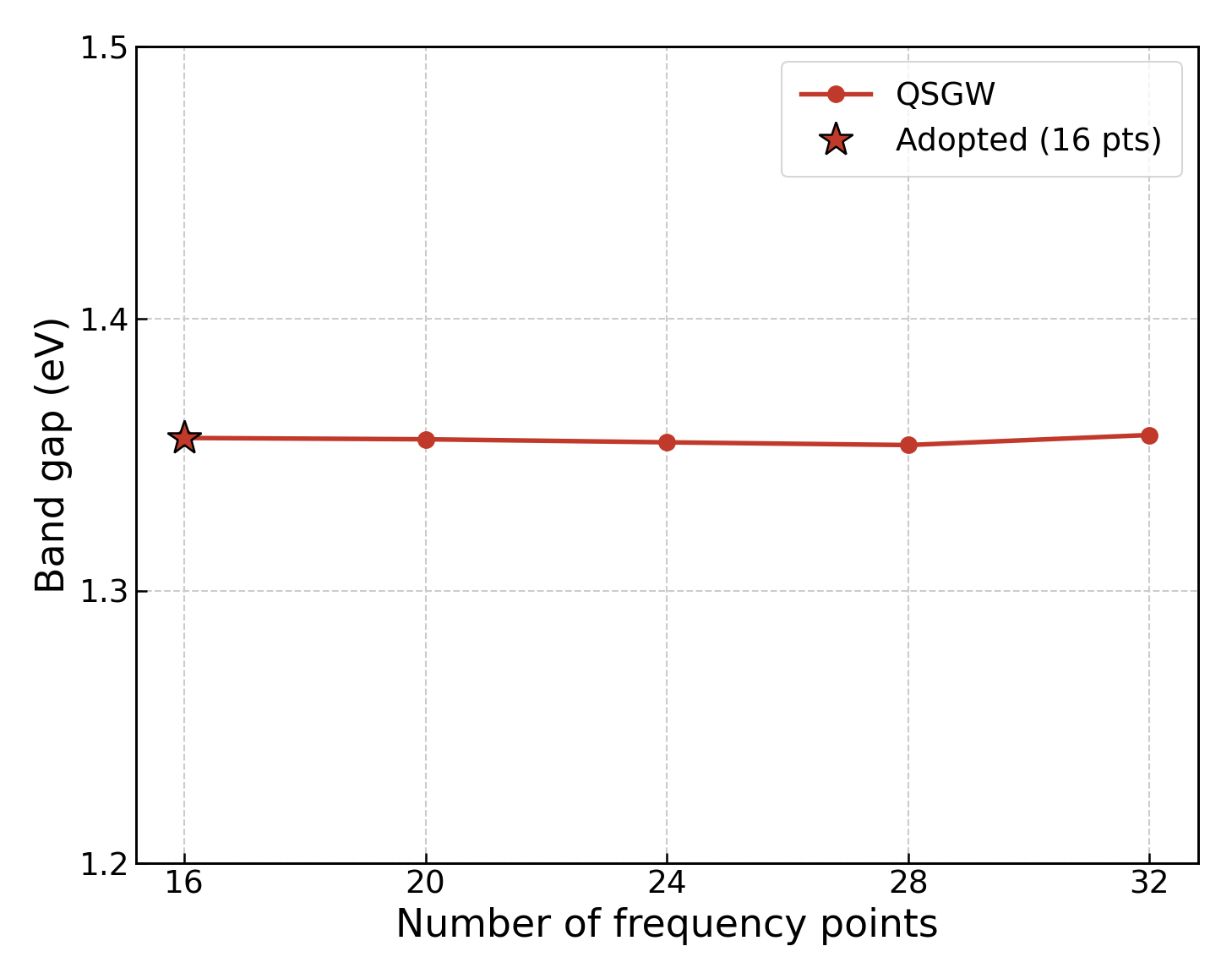}
	\caption{Frequency grid convergence of the Si QS$GW$ band gap using an $8\times8\times8$ ${\bf k}$-point mesh and the \texttt{intermediate\_gw} basis sets (see Sec.~\ref{sec:basis_set_convergence}). The calculation with 16 frequency points (highlighted with a red star) is adopted for the periodic systems in this work. The band gap varies by only 2.6 meV between 16 and 28 frequency points, confirming adequate convergence.
    }
	\label{fig:si_freq_convergence}
\end{figure}

Regarding the frequency grid, we tested the convergence of the QS$GW$ band gap with respect to the number of imaginary frequency points used in the self-energy calculation. Figure~\ref{fig:si_freq_convergence} shows the frequency grid convergence for Si using an $8\times8\times8$ ${\bf k}$-point mesh and the \texttt{intermediate\_gw} basis sets.

The results demonstrate that the QS$GW$ band gap converges rapidly with respect to the frequency grid. Between 16 and 28 frequency points, the band gap varies by only 2.6 meV (from 1.356 eV to 1.354 eV), indicating that 16 frequency points are sufficient to achieve convergence within 3 meV. We note that the calculation with 32 frequency points yields a slightly higher band gap (1.357 eV), but the difference remains at the meV level. Based on these tests, we adopt 16 frequency points for the periodic systems in this work as a practical choice that reduces the computational cost while maintaining excellent accuracy. For the molecular benchmarks reported below, we employ 32 frequency points, since the additional cost is modest compared to periodic calculations and provides a conservative numerical setting.

\subsubsection{\label{sec:basis_set_convergence} Basis sets}

For the benchmark calculations presented in this work, we adopt different basis sets for periodic solids and molecules.
For periodic systems, we employ the \texttt{intermediate\_gw} setting, which offers a practical compromise between accuracy and computational cost for the present test set.
For molecular systems, we use the \texttt{really\_tight} setting, i.e., the same orbital NAO basis as the \texttt{tight} basis sets but with more stringent numerical settings.

Here, \texttt{intermediate\_gw} and \texttt{really\_tight} denote the predefined FHI-aims \texttt{species\_defaults} setting (defaults\_2020), which explicitly specify (i) the orbital NAO basis (minimal basis plus additional confined radial functions, organized in tiers) and (ii) the associated numerical settings (integration grids and Hartree-potential accuracy) for each chemical element.
In the defaults\_2020 hierarchy, \texttt{intermediate} setting retains most numerical settings from \texttt{tight} but employs a more economical orbital basis, whereas \texttt{tight} setting corresponds to tier-based NAO orbital lists (e.g., ``tier 2'' for H--Ne and ``tier 1'' for most heavier elements), and \texttt{really\_tight} setting keeps the same orbital basis as \texttt{tight} setting but uses tightly converged numerical parameters (including \texttt{basis\_dep\_cutoff}=0).\cite{FHIaimsManual250822}


For periodic $GW$ calculations, the corresponding \texttt{intermediate\_gw} setting is identical to the standard \texttt{intermediate} setting except that it augments the ABFs used in the LRI (RI-LVL) representation of the Coulomb operator by adding extra \texttt{for\_aux} functions (specifically, \texttt{for\_aux hydro 4 f 0.0}); these \texttt{for\_aux} functions are used only to generate the auxiliary basis and do not enter the KS orbital basis.\cite{FHIaimsManual250822,Ren2021PRM}

\begin{table*}[!t]
	\centering
	\small
	\begin{tabular}{lcccc}
		\hline\hline
		Molecule   & \shortstack{Exp.\\(IP)}      & \shortstack{QS$GW$ (this work)\\(IP)} & \shortstack{QS$GW$ (ADF STO extra)\\(IP)} & \shortstack{QS$GW$ (TM def2-TZVPP)\\(IP)} \\
		\hline
		Ag$_2$     & 7.66             & 7.42  & 7.15  & ---   \\
		AlF$_3$    & 15.45*           & 16.06 & 15.72 & 15.69 \\
		As$_2$     & 10.00*           & 9.78  & 9.66  & 9.62  \\
		AsH$_3$    & 10.58*           & 10.63 & 10.70 & 10.50 \\
		BF         & 11.00            & 11.22 & 11.20 & 11.13 \\
		BH$_3$     & 12.03            & 13.65 & 13.61 & 13.52 \\
		C$_2$H$_2$ & 11.49*           & 11.52 & 11.59 & 11.53 \\
		\hline
		\textbf{MRE} (\%)  & \textemdash      & \textbf{+2.12}  & \multicolumn{2}{c}{} \\
		\textbf{MARE}  (\%)  & \textemdash      & \textbf{3.64} & \multicolumn{2}{c}{} \\
		\hline\hline
	\end{tabular}
	\caption{Experimental vertical ionization potentials (IP, in eV), together with our QS$GW$  results and independent QS$GW$ reference values from two GW100 external datasets (ADF STO extra and TM def2-TZVPP). For $GW$-based methods, IPs correspond to $-\varepsilon_\mathrm{H}^{\mathrm{QP}}$, where $\varepsilon_\mathrm{H}^{\mathrm{QP}}$ is the converged HOMO quasiparticle energy. Experimental IPs are those compiled in Ref.~\cite{vanSetten2015GW100} (Table~2 therein; values marked * are vertical, following that compilation). Each molecule employs the \texttt{really\_tight} NAO basis tier and 32 frequency points. The external QS$GW$ reference values are taken from the GW100 database~\cite{settenGW100Repo} (see main text for details). Here MARE and MRE denote the mean absolute relative error and mean relative error (in \%), respectively.
    }
	\label{tab:molecular_ip_lit}
\end{table*}

Both settings correspond to predefined FHI-aims NAO tiers and have been shown to yield well-converged $G^0W^0$ quasiparticle gaps for NAO-based all-electron implementations on similar benchmark sets (except for ZnO due to significant basis incompletion error, will be discussed below). We therefore adopt the same basis tiers for QS$GW$ to provide a consistent and well-tested numerical baseline for the present benchmarks.

\section{\label{sec:results} Benchmark results}

To comprehensively evaluate the accuracy and reliability of our NAO-based QS$GW$ implementation, we perform systematic benchmark calculations on both molecular and periodic systems.
For molecular systems, we select representative small molecules from the GW100 test set \cite{vanSetten2015GW100}.
For periodic systems, we select a diverse range of materials, including typical semiconductors, wide-gap insulators, and rare gas solids, allowing us to assess the performance of our implementation across different bonding types and band gap ranges.
In all cases, we compare our QS$GW$ results with standard DFT functionals, one-shot $G^0W^0$ calculations, other QS$GW$ implementations, and experimental values where available.

\subsection{\label{sec:molecular_benchmarks} Molecular Benchmarks}
To validate our QS$GW$ implementation for molecular systems, we perform benchmark calculations on a representative selection of small molecules from the GW100 test set \cite{vanSetten2015GW100,Kaplan2016,Caruso2016,vanSetten2013} (Table~\ref{tab:molecular_ip_lit}).
The selected molecules cover diverse bonding characters and chemical environments, including a metallic dimer (Ag$_2$), an ionic fluoride (AlF$_3$), $p$-block elemental/covalent species (As$_2$, AsH$_3$), a polar diatomic (BF), an electron-deficient hydride (BH$_3$), and a small $\pi$-bonded hydrocarbon (C$_2$H$_2$). Together, they provide a compact yet nontrivial validation set for the present implementation.

Table~\ref{tab:molecular_ip_lit} presents the vertical ionization potentials (IPs) from this work and compares them with experimental reference values and QS$GW$ benchmarks from established implementations (GW100 external datasets).
In $GW$-based quasiparticle approaches, IPs correspond to $-\varepsilon_{\mathrm{H}}^{\mathrm{QP}}$, where $\varepsilon_{\mathrm{H}}^{\mathrm{QP}}$ is the converged HOMO quasiparticle energy.

Accordingly, we compare our QS$GW$ ionization potentials against the experimental reference data reported in Ref.~\cite{vanSetten2015GW100} and against available molecular QS$GW$ benchmarks from established implementations as compiled in the GW100 database~\cite{settenGW100Repo} (Table~\ref{tab:molecular_ip_lit}). 

In Table~\ref{tab:molecular_ip_lit}, the GW100 reference datasets are labeled by the underlying code and basis settings.
In particular, ``ADF STO extra'' refers to the QS$GW$ HOMO dataset computed with the Amsterdam Density Functional (ADF) code using a Slater-type orbital (STO) basis in the larger ``extra'' basis sets~\cite{Forster2021} as provided in the GW100 database (GW100 file \texttt{qsGW\_HOMO\_ADF\_STO\_extra.json}). Likewise, ``TM def2-TZVPP'' refers to the QS$GW$ HOMO dataset computed with TURBOMOLE (TM) using the Gaussian def2-TZVPP basis sets~\cite{Caruso2016} as provided in the GW100 database (GW100 file \texttt{qsGW\_HOMO\_Tv6.0\_def2-TZVPP.json}).

This code-to-code comparison provides an additional consistency check for our molecular QS$GW$ implementation. As shown in Table~\ref{tab:molecular_ip_lit}, our QS$GW$ IPs agree well with reference QS$GW$ values from other implementations (ADF STO extra and TM def2-TZVPP), with deviations within $\sim$0.4~eV for this subset.

\subsection{\label{sec:band_gaps} Semiconductors and Insulators}
We now turn to periodic systems, where we present the QS$GW$ results for a set of semiconductors and insulators calculated by LibRPA.
The crystal structure prototypes used for the periodic benchmark set are summarized in Table~\ref{tab:solid_structures}.

\begin{table}[!t]
	\centering
	\small
	\begin{tabular*}{\columnwidth}{@{\extracolsep{\fill}}lcccc}
		\hline\hline
		System & Prototype   & $a$ (\AA) & $c$ (\AA) & $u$ \\
		\hline
		Si     & diamond     & 5.431     & \textemdash & \textemdash \\
		C      & diamond     & 3.567     & \textemdash & \textemdash \\
		SiC    & zinc blende & 4.358     & \textemdash & \textemdash \\
		BN     & zinc blende & 3.616     & \textemdash & \textemdash \\
		BP     & zinc blende & 4.538     & \textemdash & \textemdash \\
		AlP    & zinc blende & 5.463     & \textemdash & \textemdash \\
		MgO    & rock salt   & 4.211     & \textemdash & \textemdash \\
		wAlN   & wurtzite    & 3.110     & 4.980       & 0.3820 \\
		wZnO   & wurtzite    & 3.250     & 5.207       & 0.3825 \\
		GaAs   & zinc blende & 5.654     & \textemdash & \textemdash \\
		Ne     & fcc         & 4.195     & \textemdash & \textemdash \\
		\hline\hline
	\end{tabular*}
	\caption{Crystal structure prototypes and lattice constants (in \AA) used for the periodic benchmark set in Table~\ref{tab:QSGW_comparison}. For cubic systems, $c$ is not applicable. For wurtzite systems, $u$ denotes the internal fractional $z$ coordinate of the anion on the Wyckoff site $(1/3,2/3,u)$ (with the second anion at $(2/3,1/3,u+1/2)$). }
	\label{tab:solid_structures}
\end{table}

Table~\ref{tab:QSGW_comparison} presents the band gaps of a series of semiconductors and insulators computed by our NAO-based QS$GW$ implementation.
We first validate our approach against established implementations by comparing our QS$GW$ gaps to literature benchmarks from independent codes (Table~\ref{tab:QSGW_comparison}). In particular, we list QS$GW$ reference values obtained with crystalline Gaussian basis sets~\cite{Lei2022} and all-electron LAPW-based QS$GW$ benchmark values~\cite{salas2022electronic}. For additional context on the spread across widely used implementations, we also list representative self-consistent $GW$-family reference results obtained within a PMT (plane waves plus muffin-tin orbitals)  hybrid-QS$GW$ workflow~\cite{deguchi2016accurate}, a PAW-based fully self-consistent $GW$ implementation~\cite{grumet2018beyond}, and a plane-wave self-consistent $GW$ scheme with efficient vertex corrections~\cite{chen2015accurate}.

To facilitate cross-code validation, we adopt the Gaussian-QS$GW$~\cite{Lei2022} and LAPW-QS$GW$~\cite{salas2022electronic} results listed in Table~\ref{tab:QSGW_comparison} as our primary QS$GW$ reference benchmarks where available. The remaining entries—PMT hQS$GW$~\cite{deguchi2016accurate}, PAW sc$GW$~\cite{grumet2018beyond}, and PW sc$GW$+vertex~\cite{chen2015accurate}—represent closely related self-consistent $GW$-type approaches. These are included to illustrate the spread among widely used methodologies. However, since they are not strictly equivalent to the standard QS$GW$ formalism, they are treated as contextual references rather than direct benchmarks. The superscript $\mathrm{zb}$ indicates that the corresponding value for ZnO was reported for the zinc-blende structure in Ref.~\cite{grumet2018beyond}, as compiled in Ref.~\cite{Lei2022}.

\begin{table*}[!t]
	\centering
	\scriptsize
	\setlength{\tabcolsep}{2pt}
	
	\begin{tabular*}{\textwidth}{@{\extracolsep{\fill}}lcccccccccc}
		\hline\hline
		\multirow{2}{*}{System} & \multirow{2}{*}{PBE} & \multirow{2}{*}{$G^0W^0$@PBE} & \multirow{2}{*}{QS$GW$ (this work)} & \multicolumn{6}{c}{Literature (independent implementations / reference workflows)} & \multirow{2}{*}{Expt.} \\
		\cline{5-10}
		& & & & \shortstack{Gaussian\\QS$GW$~\cite{Lei2022}} & \shortstack{LAPW\\QS$GW$~\cite{salas2022electronic}} & \shortstack{PMT\\hQS$GW$~\cite{deguchi2016accurate}} & \shortstack{PAW\\sc$GW$~\cite{grumet2018beyond}} & \shortstack{PW\\sc$GW$+vertex~\cite{chen2015accurate}} & \shortstack{FP-LMTO\\QS$GW$~\cite{Kotani2007a}} & \\
		\hline

		Si & 0.59 & 1.09 & 1.36 & 1.32 & 1.39 & 1.28 & 1.49 & 1.47 & 1.25 & 1.23 \\
		C & 4.18 & 5.82 & 6.41 & 6.14 & 6.23 & 6.11 & 6.43 & 6.40 & 5.97 & 5.85 \\
		SiC & 1.38 & 2.46 & 3.17 & 2.81 & 3.00 & 2.63 & 2.88 & 2.90 & 2.53 & 2.57 \\
		BN & 4.50 & 6.42 & 7.29 & 7.07 & 7.62 & --- & 7.50 & 7.51 & --- & 6.66 \\
		BP & 1.25 & 2.01 & 2.39 & --- & --- & --- & --- & --- & --- & 2.35 \\
		AlP & 1.58 & 2.30 & 2.56 & 2.76 & --- & 2.74 & 2.94 & 3.10 & --- & 2.53 \\
		MgO & 4.74 & 7.50 & 9.14 & 9.33 & 9.97 & 8.97 & 9.58 & 9.29 & --- & 7.98 \\
		wAlN & 4.22 & 5.93 & 7.10 & 7.09 & --- & 6.91 & --- & --- & --- & 6.44 \\
		wZnO & 0.81 & 2.40 & 3.71 & 4.63 & --- & 3.88 & 4.29$^{\mathrm{zb}}$ & 4.61 & 3.87 & 3.60 \\
		GaAs & 0.53 & 1.37 & 1.94 & --- & 2.24 & --- & --- & --- & 2.15 & 1.69 \\
		Ne & 13.64 & 21.49 & 22.84 & 22.57 & --- & --- & --- & --- & --- & 21.7 \\
		\hline
		\textbf{MRE (\%)} & \textbf{-45.64} & \textbf{-10.05} & \textbf{+9.43} & \multicolumn{7}{c}{} \\
		\textbf{MARE (\%)} & \textbf{45.64} & \textbf{10.05} & \textbf{9.43} & \multicolumn{7}{c}{} \\
		\hline\hline
	\end{tabular*}
	\caption{Band gaps (in eV) of semiconductors and insulators. An $8\times8\times8$ {\bf k}-point mesh was employed for most systems, with the exception of wAlN and wZnO ($8\times8\times5$). All calculations used a frequency grid of 16 points and the \texttt{intermediate\_gw} basis sets (see Sec.~\ref{sec:basis_set_convergence}). Experimental band gaps are taken from Ref.~\cite{Ren2021PRM} and references therein, except for Ne, which is from Ref.~\cite{schwentner1975photoemission}. For BP, reported experimental values range from 2.2 to 2.5 eV; we use the midpoint value of 2.35 eV when computing error metrics. Literature reference values are cited in the column headers, with further interpretation provided in the main text. }
	\label{tab:QSGW_comparison}
\end{table*}

\begin{figure}[tbp]
	\centering
	\includegraphics[width=1\linewidth]{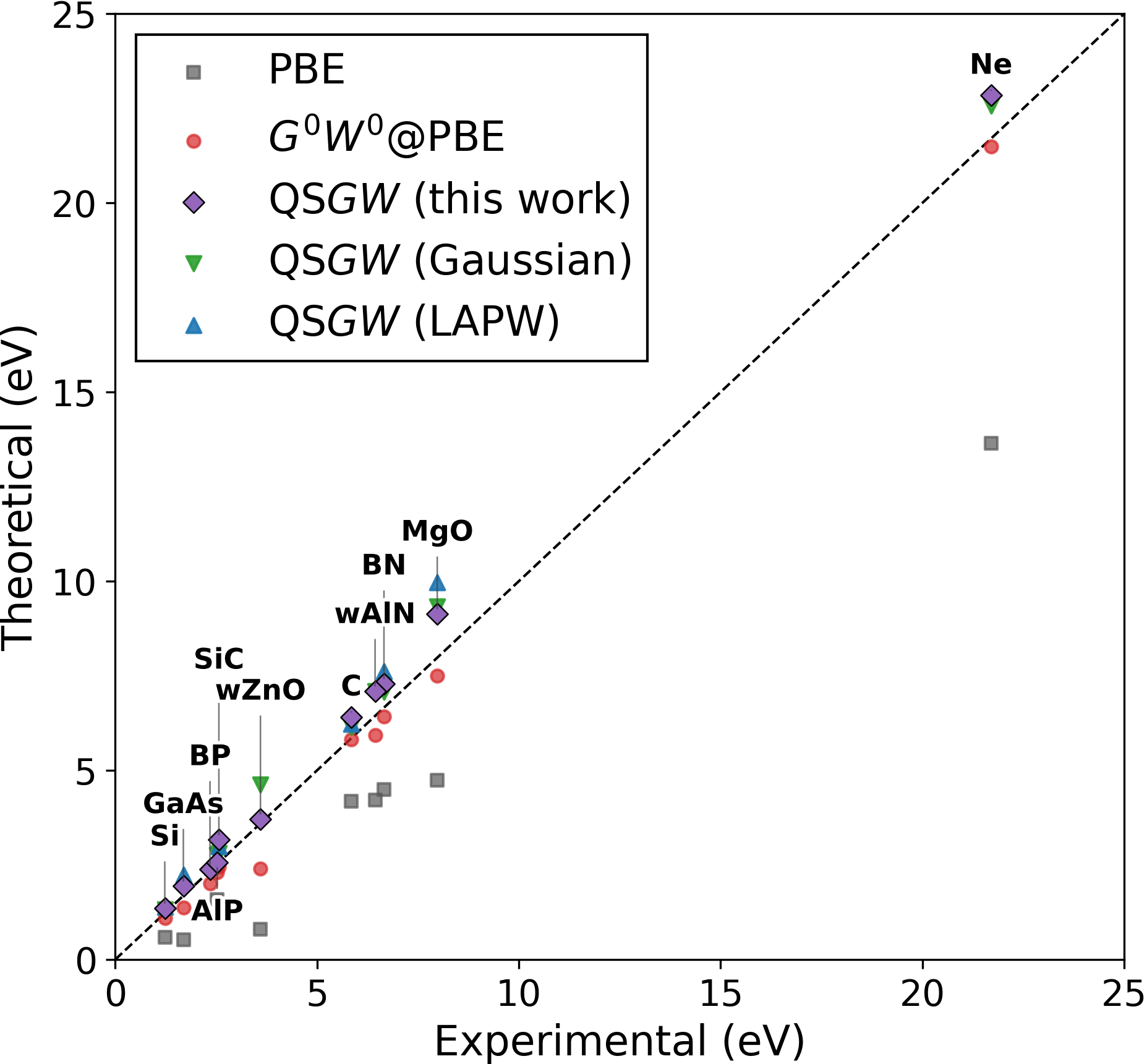}
	\caption {Band gaps of semiconductors and insulators computed with PBE, $G^0W^0$@PBE, and QS$GW$ under the numerical atomic orbital (NAO) framework in LibRPA.}
	\label{fig:band_gaps_semiconductors_insulators}
\end{figure}

As illustrated in Fig.~\ref{fig:band_gaps_semiconductors_insulators}, the PBE functional systematically underestimates the band gaps for this benchmark set. The single-shot $G^0W^0$@PBE method captures most of the gap increase relative to PBE, while QS$GW$ provides a further systematic gap opening on top of $G^0W^0$@PBE. Under the present computational protocol, QS$GW$ overestimates the band gaps by roughly the same amount as $G^0W^0$@PBE underestimates them, when compared to the experimental values.  In the following, we validate our QS$GW$ results against independent QS$GW$ benchmarks from the literature, with experimental gaps included only for contextual reference. 

\begin{figure}[tbp]
	\centering
	\includegraphics[width=\linewidth,height=0.45\textheight,keepaspectratio]{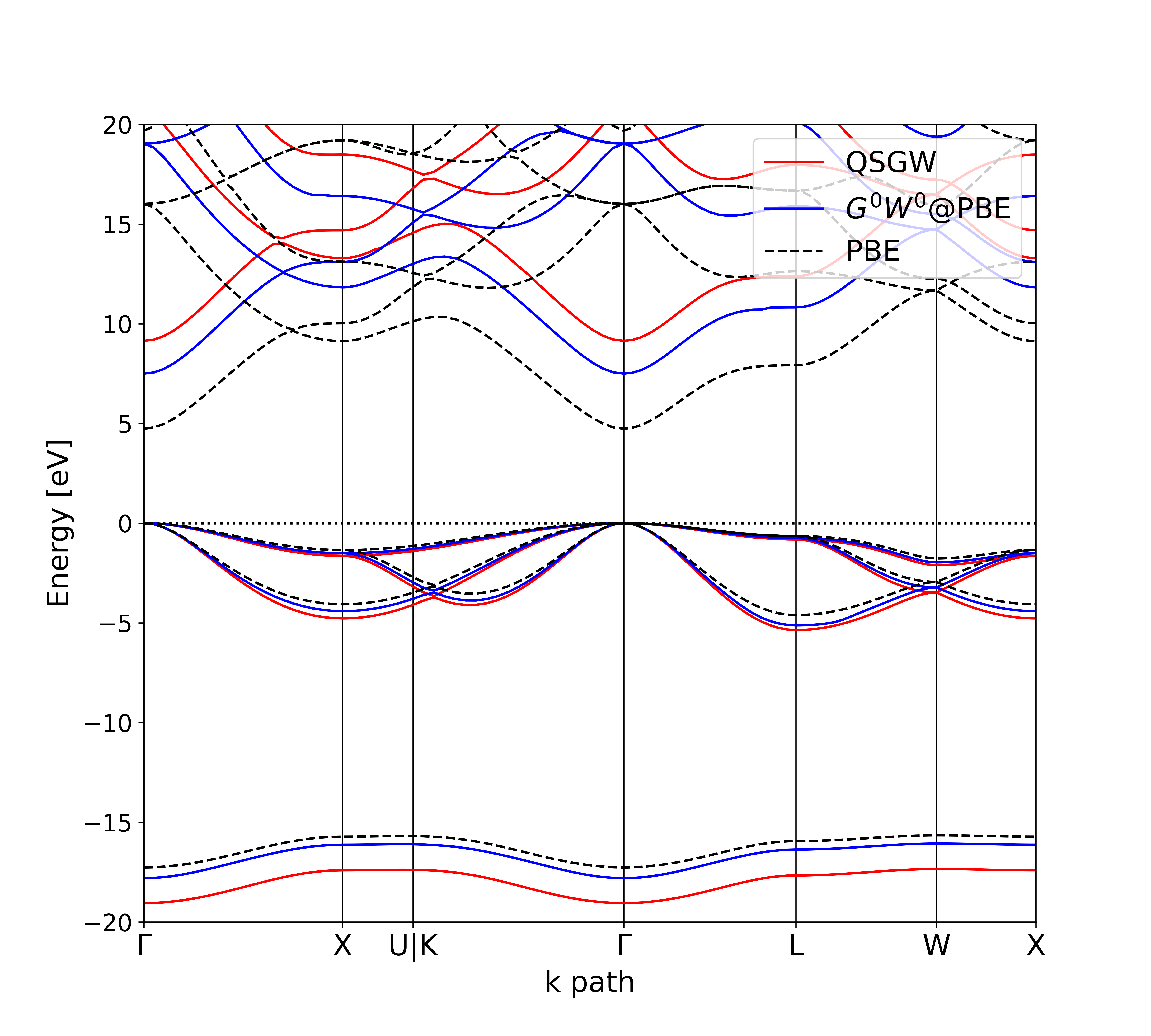}
	\caption{Band structure of MgO calculated using PBE (FHI-aims) and QS$GW$ (LibRPA). The calculations employed an $8\times8\times8$ {\bf k}-point mesh, a frequency grid of 16 points, and the \texttt{intermediate\_gw} basis sets.}
	\label{fig:MgO_band_QSGW_vs_KS}
\end{figure}

First, we assess cross-implementation consistency by comparing our QS$GW$ band gaps with literature benchmarks from independent implementations, specifically Gaussian-QS$GW$~\cite{Lei2022} and LAPW-QS$GW$~\cite{salas2022electronic} (see Table~\ref{tab:QSGW_comparison}). Across the set of materials common to both studies, our QS$GW$ results typically agree within $\sim$0.3~eV with these literature values. Especially for wide-gap insulators like C (diamond) and BN, our results (6.41~eV and 7.29~eV, respectively) are fairly close to the literature values listed in Table~\ref{tab:QSGW_comparison}. For the rare gas solid Ne, our QS$GW$ result (22.84~eV) is in excellent agreement with the periodic QS$GW$ benchmark value (22.57~eV) reported in Ref.~\cite{Lei2022} and with the experimental value (21.7~eV). Overall, the observed cross-code differences are comparable to the spread reflected in reported periodic QS$GW$ results, supporting the validity of our NAO-based QS$GW$ scheme relative to established implementations.


The largest deviations are observed for wurtzite ZnO (wZnO) (3.71~eV), which is 0.92~eV lower than the Gaussian-QS$GW$ value (4.63~eV), and for MgO (9.14~eV), which is 0.83~eV lower than the LAPW-QS$GW$ value (9.97~eV). wZnO is a system known to be challenging for the $GW$ methods, due to the strong hybridization between Zn $3d$ states and O $2p$ states, as well as the slow convergence with respect to high-energy unoccupied states \cite{FriedrichC11}.
Previous studies have shown that, in the LAPW framework, adding high-energy local orbitals (HLOs) can significantly increase the $GW$ band gap of ZnO \cite{jiang2016gw}. The underestimation of the $G^0W^0$ band
gap for ZnO within the standard NAO basis sets has been discussed in Ref.~\cite{Ren2021PRM}. It has been
shown \cite{Ren2021PRM} that the issue can be mitigated by complementing NAOs with highly localized Slater-type orbitals (STOs),
in analogy with the LAPW+HLOs scenario.


To further illustrate the quality of our QS$GW$ calculations beyond band gap values, we present the full band structure of MgO in Fig.~\ref{fig:MgO_band_QSGW_vs_KS}.
From PBE to $G^0W^0$@PBE, there is a significant increase in the band gap while preserving the overall shape of the band dispersion. From $G^0W^0$@PBE to QS$GW$, the band gap is further opened up, as expected. The smooth and continuous bands across the Brillouin zone indicate the numerical stability of our implementation under the protocol described in Secs.~\ref{QSGW_in_LibRPA}--\ref{sec:convergence}.

We further observe that the bands immediately below the valence-band maximum are very similar between QS$GW$ and $G^0W^0$@PBE, whereas deeper valence bands in the energy region around -17 eV are shifted to lower energies in QS$GW$ (Fig.~\ref{fig:MgO_band_QSGW_vs_KS}). This shift reflects the state- and energy-dependent nature of the $GW$ self-energy and the accompanying changes in screening along the self-consistent cycle, as also discussed in the literature \cite{Kotani2007a,salas2022electronic}. We expect that the energy positions of these deeper valence states are better described in QS$GW$.

To wrap up this section, benchmark calculations demonstrate that our all-electron QS$GW$ implementation in LibRPA is reliable and accurate. It exhibits the typical systematic overestimation of band gaps characteristic of the QS$GW$ approximation, consistent with other state-of-the-art implementations.
The combination of NAOs, LRI, and the space-time formalism provides a practical all-electron framework for QS$GW$ calculations for both molecules and periodic solids within LibRPA. 

\section{\label{sec:conclusion} Conclusion and outlook}

In summary, We have developed and validated an all-electron implementation of the QS$GW$ method for molecules and periodic systems within the numerical atomic orbital (NAO) framework of LibRPA, based on the space-time formalism and the LRI technique.

We analyzed numerical aspects specific to QS$GW$ in this framework, in particular, the stability of the Pad\'e analytic continuation and its basis representation dependence, and identified ``Mode B'' as the more robust construction within our current protocol.
Benchmark calculations for a set of small molecules and for representative semiconductors and insulators show that the resulting IPs and quasiparticle gaps are consistent with established reference implementations, providing strong validation of the correctness of our implementation.
For the crystalline solids, we obtain an MARE of 9.43\% (MRE of $+9.43$\%) relative to experiment, while for the molecular subset in Table~\ref{tab:molecular_ip_lit}, we obtain an MARE of 3.64\% (MRE of $+2.12$\%) for the vertical ionization potentials relative to experiment.
Consistent with the known behavior of standard QS$GW$, the gaps are systematically overestimated, reflecting under-screening of the Coulomb interaction in the absence of vertex corrections \cite{VanSchilfgaarde2006,Bruneval2014}.

Looking forward, the present NAO-based QS$GW$ framework provides a practical starting point for incorporating vertex corrections in $W$ and/or $\Sigma$ \cite{shishkin2007accurate,Cunningham2023,Kutepov2022,Forster2022_W} and for combining QS$GW$ with dynamical mean-field theory for strongly correlated materials \cite{Biermann2003,Kotliar2006,Choi2016,Zhu2021_DMFT,Lee2017,Nilsson2017}.
On the numerical side, more robust analytic continuation schemes and further basis-set optimization will be important to reduce residual instabilities in challenging cases.
In particular, addressing basis-set completeness for ionic oxides (e.g., ZnO) via targeted basis augmentation and systematic extrapolation is an important next step.
Finally, the availability of a quasiparticle Hamiltonian enables extensions to neutral excitations via QS$GW$--BSE \cite{Forster2022_BSE} and to other applications such as core-level binding energies \cite{Li2022}, relativistic two-component $GW$ \cite{Forster2023}, magnetic systems \cite{Pokhilko2022}, and data-driven modeling \cite{takano2025machine}.
Going beyond the $GW$ approximation to include higher-order correlations \cite{Cunningham2024} remains an active research direction.

Last but not least, our present QS$GW$ is built upon the existing low-scaling $G^0W^0$ implementation in LibRPA, which enables efficient $G^0W^0$ calculations for systems containing up to $O(10^3)$ atoms \cite{Zhang2026arXiv}. This low-scaling algorithm directly carries over to QS$GW$, and benchmark calculations for its performance on large-scale systems are ongoing. We expect that our QS$GW$ implementation will provide a highly useful tool for investigating complex materials systems. 

\section*{\label{sec:data} Data availability}
The data that support the findings of this study are available upon reasonable request from the authors.

\begin{acknowledgments}
	We acknowledge the funding support from the National Key Research and Development Program of China (Grant Nos.2023YFA1507004 and 2022YFA1403800) and the Strategic Priority Research Program of the Chinese Academy of Sciences under Grant No. XDB0500201. This work was also supported by the National Natural Science Foundation of China (Grants Nos. 12134012, 12374067, and 12188101).
\end{acknowledgments}

\bibliography{./CommonBib}

\end{document}